\documentclass[preprint]{aa}
\usepackage{rotating}   
\usepackage{graphicx}
\usepackage{txfonts}
\usepackage{natbib}   

\newcommand{\percc}{cm$^{-3} $}        
\newcommand{\persc}{cm$^{-2} $}        
\newcommand{\kks}{K km s$^{-1} $}      
\newcommand{\kms}{km s$^{-1} $}        
\newcommand{\HH}{H$_2 $}               
\newcommand{\HCOP}{HCO$^+ $}           
\newcommand{\HTHCOP}{H$^{13}$CO$^+ $}  
\newcommand{\TWCO}{$^{12}$CO}          
\newcommand{\THCO}{$^{13}$CO}          
\newcommand{\CEIO}{C$^{18}$O$ $}       
\newcommand{\CSEO}{C$^{17}$O$ $}       
\newcommand{\NTHP}{N$_2$H$^+ $}        
\newcommand{\NTDP}{N$_2$D$^+ $}        
\newcommand{\DoH}{$N$(N$_2$D$^+$)/$N$(N$_2$H$^+$)}

\begin{document}

\title{Dynamical and chemical  properties of the ``starless'' core L1014}

\author{Antonio Crapsi\inst{1,2}  \and
	Christopher H. DeVries\inst{1} \and
	Tracy L. Huard\inst{1} \and
	Jeong-Eun Lee\inst{3} \and
	Philip C. Myers\inst{1}  \and
	Naomi A. Ridge\inst{1} \and
        Tyler L. Bourke\inst{1,4} \and
	Neal J. Evans II\inst{3} \and
	Jes K. J{\o}rgensen\inst{1,5} \and
	Jens Kauffmann\inst{6} \and
	Chang Won Lee\inst{7} \and
	Yancy L. Shirley\inst{8} \and
	Chadwick H., Young\inst{3}}
   \offprints{A. Crapsi, \email{crapsi@arcetri.astro.it}}

\institute{
 Harvard--Smithsonian Center for Astrophysics, 60 Garden Street, Cambridge, MA 02138, USA
   \and
 Universit\`a degli Studi di Firenze, Dipartimento di Astronomia e Scienza dello Spazio, 
 	  Largo E. Fermi 5, I-50125 Firenze, Italy
   \and
 University of Texas at Austin, 1 University Station C1400, Austin, TX 78712-0259
   \and
 Harvard--Smithsonian Center for Astrophysics, Submillimeter Array Project, 645 N.
         A'ohoku Place, Hilo, HI 96720, USA 
   \and
 Leiden Observatory, P.O. Box 9513, 2300 RA Leiden, Netherlands
   \and
 Max-Planck-Institut F\"{u}r Radioastronomie (MPIfR), Bonn, Germany
   \and
 Korea Astronomy and Space Science Institute, 61-1 Hwaam-dong, Yusung-gu, Daejon 305-348, Korea
   \and
 National Radio Astronomical Observatory, P.O. Box 0, Socorro, NM 87801}

   \date{Received 29/11/2004 / Accepted 02/02/2005}

\titlerunning{L1014 molecular observations}
\authorrunning{Crapsi et al.}

\abstract{
 Spitzer Space Telescope observations of a point-like source, L1014-IRS,
 close to the dust peak of the low-mass dense core L1014 have questioned its
 starless nature.  The presence of an object with colors expected for an
 embedded protostar makes L1014-IRS the lowest luminosity isolated
 protostar known, and an ideal target with which to test star formation
 theories at the low mass end.
 
 In order to study its molecular content and to search for the presence of a molecular outflow,
 we mapped L1014 in at least one transition of \TWCO, \NTHP, \HCOP,  CS and of their 
  isotopologues \THCO,  \CEIO,  \CSEO, \NTDP and \HTHCOP,
 using the Five College Radio Astronomy Observatory (FCRAO),
 the IRAM 30 meter antenna and the Caltech Submillimeter Observatory (CSO).
 
 The data show physical and chemical properties in L1014 typical of the less evolved starless cores: 
 i.e. \HH \ central density of a few 10$^5$~molecules~\percc , estimated  mass of $\sim$2~M$_\odot$,
 CO integrated depletion factor less than 10,  $N$(\NTHP)$\simeq 6 \times 10^{12}$~\persc , 
 \DoH \ equal to 10\% and relatively broad \NTHP(1--0) lines (0.35~\kms). 
 Infall signatures and significant velocity shifts between
 optically thick and optically thin tracers are not observed in the line profiles.

 No classical signatures of molecular outflow are found in the \TWCO \ and \THCO \ observations. 
 In particular, no high velocity wings are found, and no well-defined blue-red lobes of \TWCO \ emission are 
 seen in the channel maps. 
 If sensitive, higher resolution observations confirm the absence of an outflow on a smaller scale than
 probed by our observations, L1014-IRS would be the only protostellar object known to be formed 
 without driving an outflow.
 
\keywords{ISM: clouds --  ISM: evolution  -- ISM: individual(L1014)  --
          ISM: molecules --  ISM: jets and outflows -- Stars: formation}

 }

\maketitle

\section{Introduction}
  Starless cores are cold ($\sim$ 10-20~K) and dense ($> 10^4$~\percc) 
  condensations of gas and dust in which no sign of the presence of a central 
  protostellar object has been found. Previously, the easiest way to determine 
  if dust emission was associated with a protostellar object was to search for a source 
  emitting at mid-infrared wavelengths in the $IRAS$ catalogue  \citep[see, ][]{beichman1986}. 
  This technique was obviously limited by the  $IRAS$ sensitivity. 
  
  A clear example is represented by \object{IRAM 04191}: 
  although \citet{andre1999} found a clear sign of the presence of a Class~0 
  object represented by  a clear collimated bipolar outflow departing from the core  peak, 
  $IRAS$ does not show a point source towards the dust emission peak. 
  Moreover, the CS line they observed towards the nucleus confirmed the presence of star formation
  activity showing a clear double peaked profile with the blue peak brighter than the red one. 
  This spectral asymmetry has been recognized as an indicator of systematic inward motion 
  \citep{snell1977, zhou1992, tafalla1998, lee1999}. 

  In the same fashion, we present here the case of \object{L1014}. 
  This core, listed as  an opacity class 6 object in the \citet{lynds1962} catalogue, 
  lies $-$0\fdg25  below the galactic plane, has a line of sight velocity of $\sim$4~\kms \ 
  with respect to the local standard of rest  and is projected just 10\arcmin \ south of another 
  dark globule  B362  \footnote{ We note that B362 and L1014 were called  ``L1014-1'' 
  and ``L1014-2'' in \citet{lee1999a}. In the present paper we keep the original names as in  
  \citet{barnard1927} and \citet{lynds1962}.  }.
  These cores can be seen in the Digital Sky Survey optical
  image presented in Figure~\ref{Fopt}, where we have overlaid visual
  extinction contours derived from near-infrared color excesses
  of background stars listed in the 2MASS catalog using
  the {\it NICE} technique \citep[e.g.][]{lada1994, alves1998}.\\
  L1014 was  included in a survey for infall asymmetry in starless cores performed
  observing CS(2--1)  and \NTHP(1--0)   \citep{lee1999}; given the weak detection in CS, no 
  \NTHP \ observations were attempted.
  The dust continuum from L1014 was first detected  at 260~$\mu$m using the NASA Kuiper Airborne 
  Observatory (KAO) \citet{keene1981}.
  Subsequently a 850~$\mu$m emission map \citep{visser2001} taken with SCUBA showed a dust emission 
  peak at (21$^{\rm h}$24$^{\rm m}$07\fs6, 49\degr59\arcmin02\arcsec, J2000) coincident with
  the visual extinction peak. 
  These authors then performed an unsuccessful search for high velocity gas, observing \TWCO(2--1) 
  at five points around the dust peak with a sensitivity of 0.3~\kks \  in a 0.2~\kms channel.

  The Spitzer Space Telescope (hereafter Spitzer) observed L1014 in December
  2003 as part of the Legacy program "From Molecular Cores to Planet Forming Disks" \citep{evans2003}.  
  Surprisingly, a strong (81.8$\pm$16~mJy)  point  source was detected  with MIPS at 24~$\mu$m towards 
  the center of L1014 \citep{young2004}, being coincident with both the 850~$\mu$m
  peak \citep{visser2001} and 1.2~mm peak (\citealt{young2004}; Kauffmann et al., in preparation). 
   This object, referred to as \object{L1014-IRS}, is also clearly detected in each of the 4 IRAC bands
  (3.6, 4.5, 5.8 and 8.0~$\mu$m) and at 70~$\mu$m with MIPS and, most importantly, shows a 
  Spectral Energy Distribution (SED) compatible with an embedded protostar having an  effective 
  temperature of $700 \pm 300$~K according to the IRAC data at $\lambda \leq 8 ~\mu$m.
   Considering that L1014-IRS is the only source visible at 70~$\mu$m in the 5\arcmin$\times$5\arcmin \
  field observed with the Spitzer and L1014 is the only starless core present in the same region,
  \citet{young2004} concluded that a chance alignment of the two sources was unlikely although 
  not negligible.
  Naively, dividing  the Spitzer field of view in squared cells with 30\arcsec \ of side, the 
  chance probability  of having both L1014 and L1014-IRS in the same cell would be 1\%.
  In this paper, although we will consider  the chance alignment as an alternative explanation,
  we will devote more discussion to the hypothesis of association 
  since the weight of the evidence supports  this 
  hypothesis.  
  Assuming a distance to the core of 200~pc, \citet{young2004} determine a luminosity of 
  0.09~L$_\odot$  for  L1014-IRS.
  From the bolometric temperature ($T_{bol}$= 50~K) and the ratio between bolometric 
  and sub-millimeter luminosity ($L_{bol}/L_{smm}$= 20), \citet{young2004} classified
  L1014-IRS as a Class~0 protostar \citep[see][for a review on Class~0 properties]{andre1994}.

  Outflows are an ubiquitous tracer of protostellar activity \citep{andre1994,richer2000}, 
  and so, in an attempt to support the classification of L1014-IRS as a
  Class~0 protostar, we deepened the search for high velocity \TWCO \ with a better 
  combination of  resolution and sensitivity than in past searches. 
  We report here on observations of \TWCO, \THCO, \CEIO, \CSEO, CS, \NTHP, \NTDP, \HCOP \ and \HTHCOP \
  aimed at surveying the molecular content of L1014 and at studying its kinematical properties.

\section{Observations}\label{obs}
  Observations were performed during March and April 2004, using the 32-pixel SEQUOIA focal
  plane array mounted on the Five College Radio Astronomy Observatory (FCRAO) 14 meter telescope.
  \\
  We mapped L1014 in CO(1--0), \THCO(1--0), \CEIO(1--0), \NTHP(1--0) and CS(2--1) using the on-the-fly 
  position  switching  mode.
  Adopted frequencies, telescope half power beam widths, system temperatures, channel spacings and size 
  of the mapped area are in Table~\ref{Tfreq}. Temperatures were converted in the main beam 
  brightness scale according to the efficiencies  tabulated in the FCRAO web 
  page\footnote{http://www-astro.phast.umass.edu/$\sim$fcrao/observer/status14m.html\#ANTENNA}.  
  Data were reduced and convolved onto a 20\arcsec \ spaced grid (circa Nyquist sampling) using the 
  otftool software 
  \citep{heyer2001}\footnote{http://www-astro.phast.umass.edu/$\sim$fcrao/library/manuals/otfmanual.html}  
  available at FCRAO  and then analysed using CLASS, the spectral line analysis 
  software of IRAM and  Observatoire de Grenoble \citep[see][for a manual]{buisson2002}.\\
  Angular offsets throughout this paper are relative to the 1.2-mm continuum peak position 
  (21$^{\rm h}$24$^{\rm m}$07\fs5, 49\degr59\arcmin05\arcsec, J2000), which is 4\arcsec \ south 
  of L1014-IRS position. 
  
  Complementary higher-resolution observations of \NTHP(1--0), (3--2), \NTDP(1--0), (2--1), (3--2) 
  and \CSEO(1--0)  were obtained in August 2004 using the IRAM-30m telescope.
  \\
  The observations consisted in a 5-pointings cross spaced by 20\arcsec \ around the 1.2-mm peak 
  position for  \NTHP(1--0), \NTDP(1--0) and \NTDP(3--2), while \NTHP(3--2) and \NTDP(2--1) were
  observed only at the (0,0), (20,0) and ($-$20,0) and \CSEO(1--0) only at the peak position.
  These data were taken in frequency switching mode using the facility autocorrelator as the backend.
  Information about the telescope parameters (channel spacing, HPBW, system temperatures) are reported 
  in Table~\ref{Tfreq2}.
  Intensities were converted in the main beam brightness scale using the efficiencies reported in the 
  IRAM web site\footnote{http://www.iram.fr/IRAMES/telescope/telescopeSummary/telescope\_summary.html}.
  
  Finally, we used the Caltech Submillimeter Observatory in August 2004 to obtain  
  \HCOP(3--2),  \HTHCOP(3--2) and  HCN(3--2) spectra towards the central position and a small 
  map in \CEIO(2--1) and \CSEO(2--1). Spectra were obtained in position switching mode. 
  Although the single pointing observations were performed 6\arcsec \ south of L1014-IRS, 
  the $\sim$30\arcsec \ beam allows us to consider them as relative to the infrared source position.
  Conversion factor to the main beam temperature ($\eta_{_{\rm 230GHz}}=0.64$) was evaluated  
  from observations of planets.
  Other informations for the telescope set-up are reported in Table~\ref{Tfreq2}.

  The pointing accuracy for all our observations was measured to be $\approx 4-6$\arcsec.

\section{Results}\label{res}
 \subsection{Maps}
\begin{figure}[htbp]
 \resizebox{12cm}{!}{\includegraphics{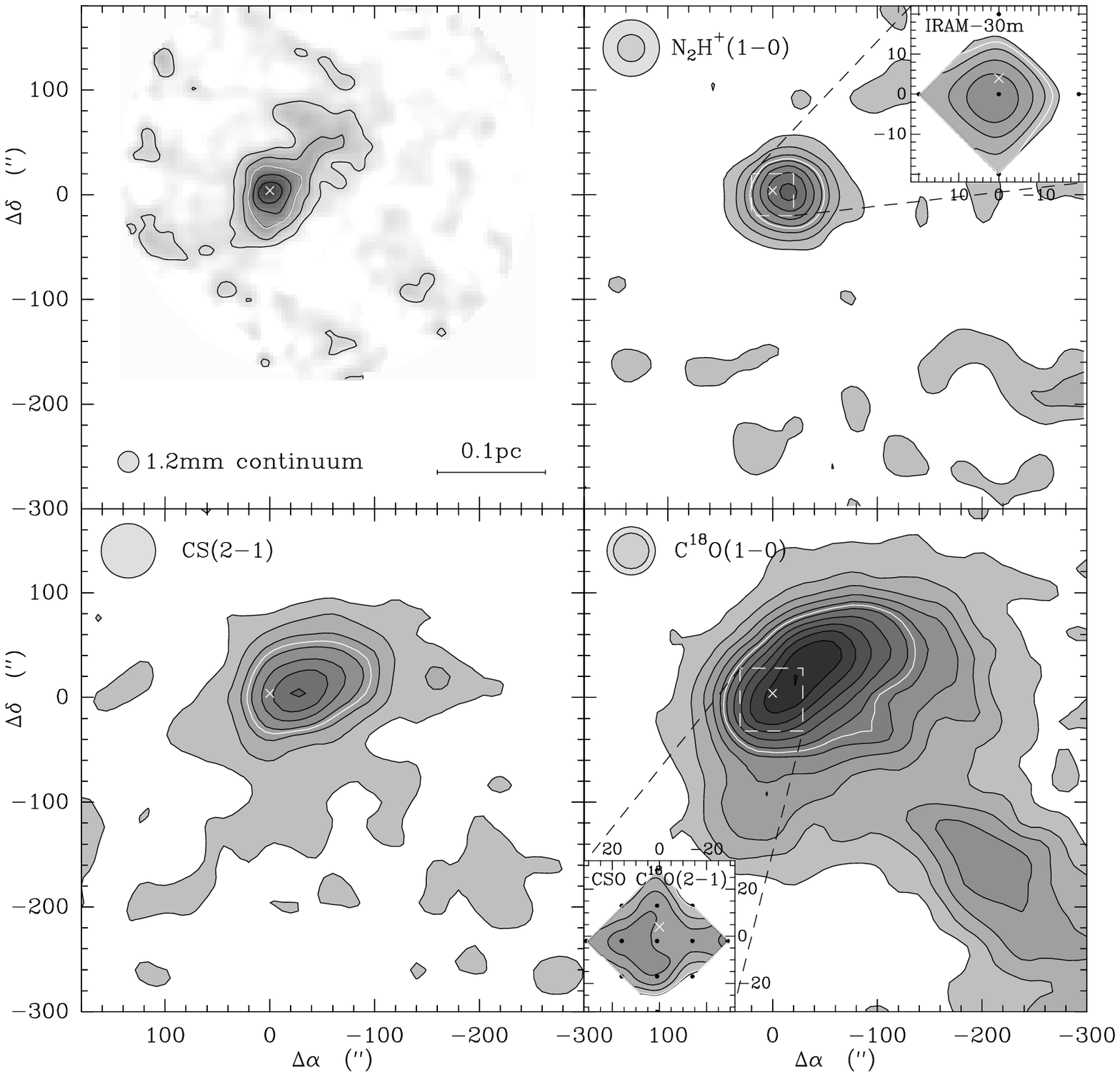}}
 \caption{1.2-mm continuum, \NTHP(1--0), CS(2--1) and \CEIO(1--0) emission towards L1014.
 Molecular line maps were taken at FCRAO with the exception of the \NTHP(1--0) map
 in the inset (from IRAM-30m) and the \CEIO(2--1) map in the inset (from CSO)
 The 1.2-mm continuum was taken from \citet{young2004} and it was observed with IRAM-30m. 
 Contour levels start at and increase by 0.08~\kks \ for the FCRAO molecular data and
 4~mJy/(11\arcsec \ beam)$^{-1}$ for the dust continuum . The white contour represents 
 the half  peak intensity in all the maps. Beam sizes are displayed at the top-left of each
 map but the 1.2-mm map. 
 The white cross places the position of a point like source detected by the Spitzer Space 
 Telescope with colors compatible with an embedded protostar. Angular offsets are relative to the 
 1.2-mm continuum peak (21$^{\rm h}$24$^{\rm m}$07\fs5, 49\degr59\arcmin05\arcsec, J2000). 
 The inset on the \NTHP(1--0) FCRAO map shows the 5 points observations of \NTHP(1--0) at IRAM-30m.
 Levels start at 0.8~\kks and increase by  0.3~\kks . The IRAM-30m beam size is shown inside that of 
 the FCRAO. 
 Similarly, the inset in the \CEIO(1--0) FCRAO map shows the 13 points map of \CEIO(2--1) taken at CSO.
 Levels start at 0.35~\kks and increase by  0.1~\kks . The CSO beam size is shown inside that of 
 the FCRAO.  
 \label{Fmaps}}
\end{figure}

  In Figure~\ref{Fmaps}, we present integrated intensity emission maps of L1014 
  (V$_{\rm LSR}\simeq 4.2$~\kms) in \NTHP(1--0), CS(2--1) 
  and \CEIO(1--0) taken with FCRAO together with the 1.2-mm continuum from \citet{young2004},
  the 5-points map of \NTHP(1--0) taken with the IRAM-30m and the 13-points map of \CEIO(2--1)
  from CSO. 
  The half maximum contour of each map is rendered in white. The white cross shows the position of
  L1014-IRS, the candidate protostellar object embedded in L1014 according to \citet{young2004}. 
  The area observed with the FCRAO is 10\arcmin$\times$10\arcmin \ in size, but only the inner 
  8\arcmin$\times$8\arcmin \ region where  emission was detected is shown in Figure~\ref{Fmaps}.

  \NTHP(1--0) emission, which comes from the high density nucleus of the core, is very concentrated
  (FWHM $\sim$70\arcsec, in the FCRAO observations) and peaks $\sim$10\arcsec \ west of the dust emission 
  (within the uncertainties due to the 54\arcsec \ FCRAO beam width  and the Nyquist sampling of the 
  observations).  
  Since the emitting region is comparable in size with the telescope beam some dilution is very likely. 
  In fact, IRAM-30m observations of \NTHP(1--0) with a 26\arcsec \ beam showed spectra 3 times brighter 
  than those observed by FCRAO towards the dust continuum peak, although still the emission does not
  seem to be fully resolved. 
  In these higher resolution observations, the integrated intensity peak falls 5\arcsec \ south  of
  L1014-IRS.
  We note that the equivalent radius of the \NTHP(1--0) emission from FCRAO ($=$0.034~pc, evaluated as 
  the square root of the area within the 50\% contour divided by $\pi$) is among the smallest
  radii found in \citet{lee2001} and \citet{caselli2002c}, which observed 63
  starless cores altogether using FCRAO and evaluated the equivalent radius with the same technique. 
  The integrated intensity of \NTHP(1--0) at peak observed with FCRAO is also lower  (a factor 
  of 2) than  any other core reported in \citet{lee2001} and \citet{caselli2002c}, although the beam 
  dilution should be kept in mind.
  Following \citet{caselli2002c}, we calculated the virial mass of L1014 from
  ${\rm M_{vir}(M_\odot)}=210 \, r  \, ({\rm pc}) \,  \Delta v_m^2  \, ({\rm km^2 s^{-2}})$, 
  where $r$ is the typical radius of the \NTHP \ emission from FCRAO and $\Delta v_m^2$ is the velocity 
  dispersion of the  mean mass molecule \citep[see][for the definition]{caselli2002c}. In this way we
  obtained a virial mass of 2.1~M$_\odot$, again, among the smallest in the \citet{caselli2002c} sample.
  This estimate is in agreement with the mass evaluated from the extinction map in Figure~\ref{Fopt}.
  In fact, using a conversion factor of $1 \times 10^{21}\; {\rm molecules \;cm}^{-2}$ per magnitude of visual 
  extinction \citep{bohlin1978}, we obtained a mass of 1.5~M$_\odot$ above the A$_V=5$ magnitudes contour.

  The CS(2--1) emission is much more extended than the 1.2-mm continuum and \NTHP(1--0), as expected
  due to its lower critical density and possible depletion effects \citep{tafalla2002}. 
  The integrated intensity peak lies $\sim$30\arcsec \ west of the dust emission, although the same
  caveats about the limited resolution and sampling apply here as in the \NTHP \ map.

  \CEIO(1--0) observations reveal more structure than previous observations, showing a secondary 
  peak at ($-$200\arcsec,$-$200\arcsec) from the reference position. Note that the 1.2-mm map did not cover 
  that area. Also the \CEIO(1--0) integrated intensity peaks west of the continuum, although
  the peak is less constrained than the other tracers.
  The flatness of the \CEIO \ map and the offset of its peak with respect to the dust are typical 
  features of starless cores \citep{tafalla2002}, where CO (and its isotopologues) freeze-out onto 
  dust grains in the high density nucleus. To test if this is the situation in L1014, we 
  follow \citet{caselli2002b} evaluating the degree
  of CO depletion as the ratio of the canonical CO abundance ([CO]/[\HH]$\equiv 9.5 \times 10^{-5}$, 
  \citealt{frerking1982}) and the observed abundance derived from the ratio of \CEIO \ and \HH \ column 
  densities \citep[for details see][]{caselli2002b}. 
  Following \citet{crapsi2005}, the CO column density was derived under the constant excitation temperature 
  approximation  \citep[CTEX, see eq. A4 in][]{caselli2002b}  assuming that \CEIO \ emission is
  optically thin.  In this way we derived a  \CEIO \ column density of $7.8\times 10^{14}$~\persc \ at 
  the dust peak  position which increases to  $8.2\times 10^{14}$~\persc \ at the \CEIO \ peak
  position ($-$20\arcsec, 20\arcsec). The CO column density can then be inferred  from the local 
  interstellar medium relative abundance value  ([CO]/[\CEIO]=560; \citealt{wilson1994}).
  \HH \ column density was inferred from the 1.2-mm emission assuming constant dust temperature (10~K) and 
  emissivity ($\kappa_{1.2mm}=0.005$~cm$^2$~g$^{-1}$)  similarly to \citet{wardthompson1999}, obtaining
  $ N(H_2) = 4\times 10^{22}$~\persc \ at the dust peak position and $ N(H_2) = 2.2\times 10^{22}$~\persc \ 
  at the \CEIO \ peak position.
  We thus found a depletion factor around 9 at the dust  peak, whereas it drops to 5 at the \CEIO \ peak.
  These values are typical of starless cores with moderate chemical evolution (e.g. L1495, L1498, L492) 
  with densities of about a few $10^5$~\percc \citep{crapsi2005}.\\
  The 30\arcsec \ map of \CEIO(2--1) obtained at CSO is shown in an inset of Figure~\ref{Fmaps}. 
  Note that the error on the integrated intensity for the points outside the peak is 2.5 times the
  value reported in Table~\ref{Tint}.
  Similarly to the \NTHP \ high--resolution observations, the \CEIO(2--1) from CSO seems to peak south of 
  L1014-IRS.

  While the 1.2-mm emission peak in L1014 (23~mJy~(11\arcsec \ beam)$^{-1}$) is comparable in intensity 
  with the majority of \NTHP \ emitting starless cores \citep[see e.g][]{crapsi2005}, it is relatively 
  weak if  compared with  {\it high-density, evolved} pre--stellar cores such as L1544 
  ($>$ 60~mJy~(11\arcsec \ beam)$^{-1}$, \citealt{wardthompson1999}). 
  Consequently, the \HH \ central volume density inferred from it is 6 times 
  less ($2.5 \times 10^5$~\percc \ in L1014 vs. $1.4 \times 10^6$~\percc \ in L1544)
  when evaluated with the same technique and physical parameters as in \citet{tafalla2002}.
   This technique derives the density profile of a spherically symmetric core  which best fit the
  observed 1.2-mm continuum. In our calculations we adopted a density profile model of the form 
  $n({\rm H_2})=n_0/(1+r/r_0)^\alpha$, finding, in addition to the central density, the radius of the 
  ``flat'' region  $\sim$20\arcsec \ and the asymptotic power law index $\sim$2.7.
  Note that the central \HH \ density value given in \citet{young2004} ($1.5 \times 10^5$~\percc )
  was evaluated using a dust opacity equal to 0.0104~cm$^2$~g$^{-1}$.

 \subsection{Spectra towards dust peak}\label{spec}
\begin{figure}[htbp]
 \begin{center}
 \resizebox{12cm}{!}{\includegraphics{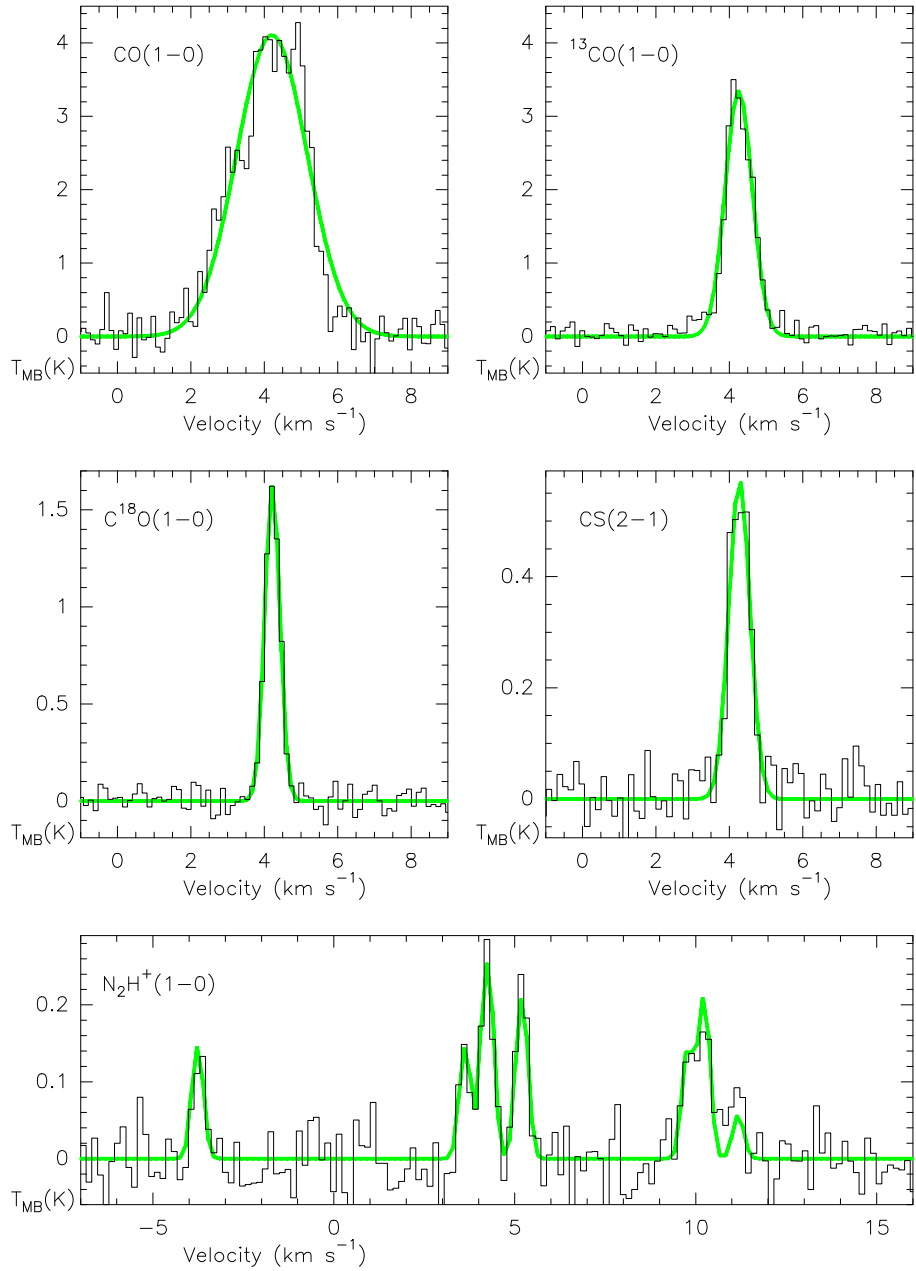}}
 \caption{CO(1--0), \THCO(1--0), \CEIO(1--0), CS(2--1) and  \NTHP(1--0) spectra observed 
 with FCRAO towards the 1.2-mm peak of L1014. Gaussian (or hyperfine in the case of \NTHP(1--0)) 
 fits are plotted. \label{Fspec}}
 \end{center}
\end{figure}

\begin{figure}[htbp]
 \begin{center}
 \resizebox{10cm}{!}{\includegraphics{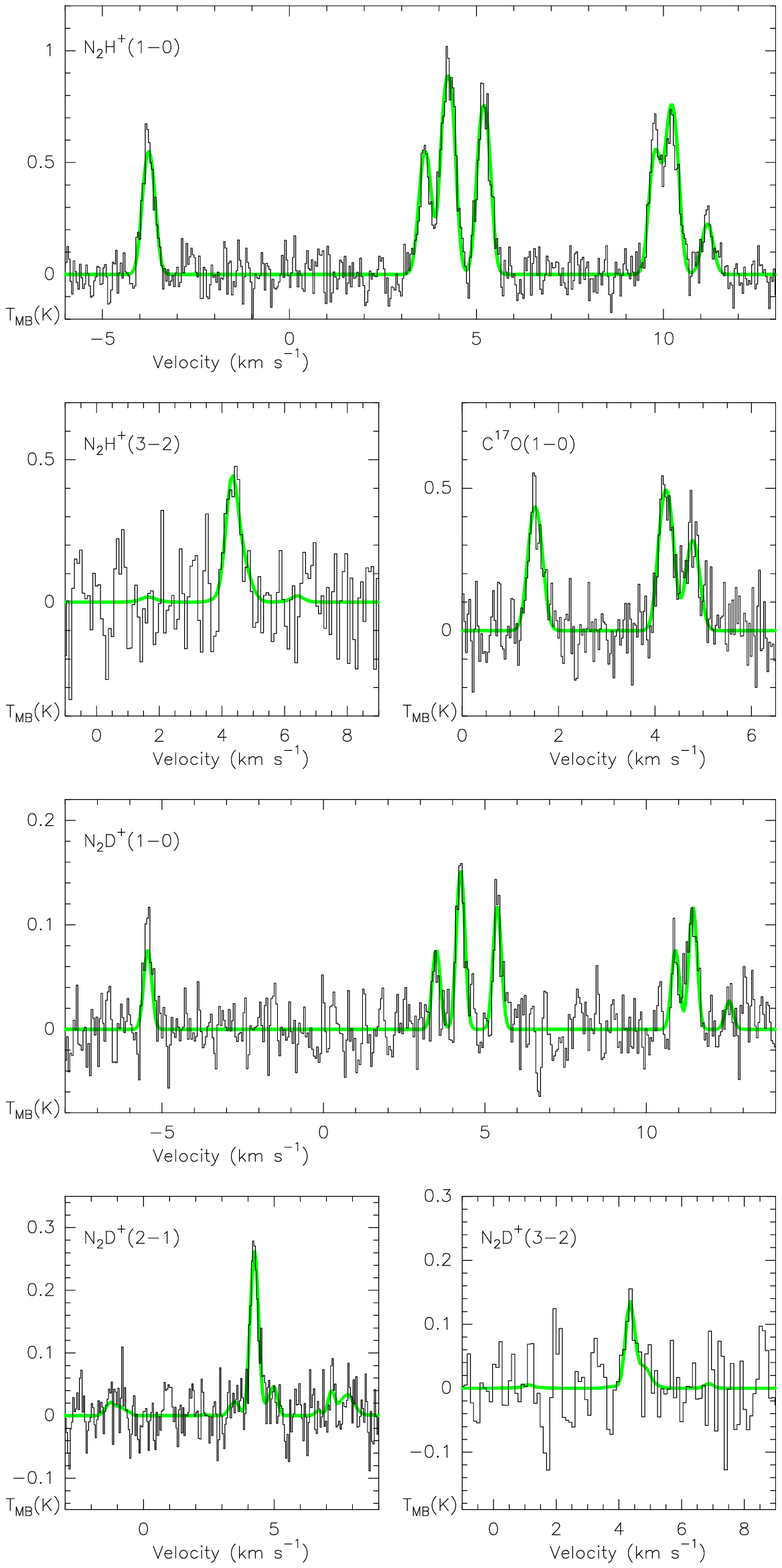}}
 \caption{\NTHP(1--0), \NTHP(3--2), \CSEO(1--0), \NTDP(1--0), \NTDP(2--1) and \NTDP(3--2) spectra observed 
 with the IRAM-30m towards the 1.2-mm peak of L1014. Hyperfine fits are plotted. \label{Fspec2}}
 \end{center}
\end{figure}

\begin{figure}[htbp]
 \begin{center}
 \resizebox{10cm}{!}{\includegraphics{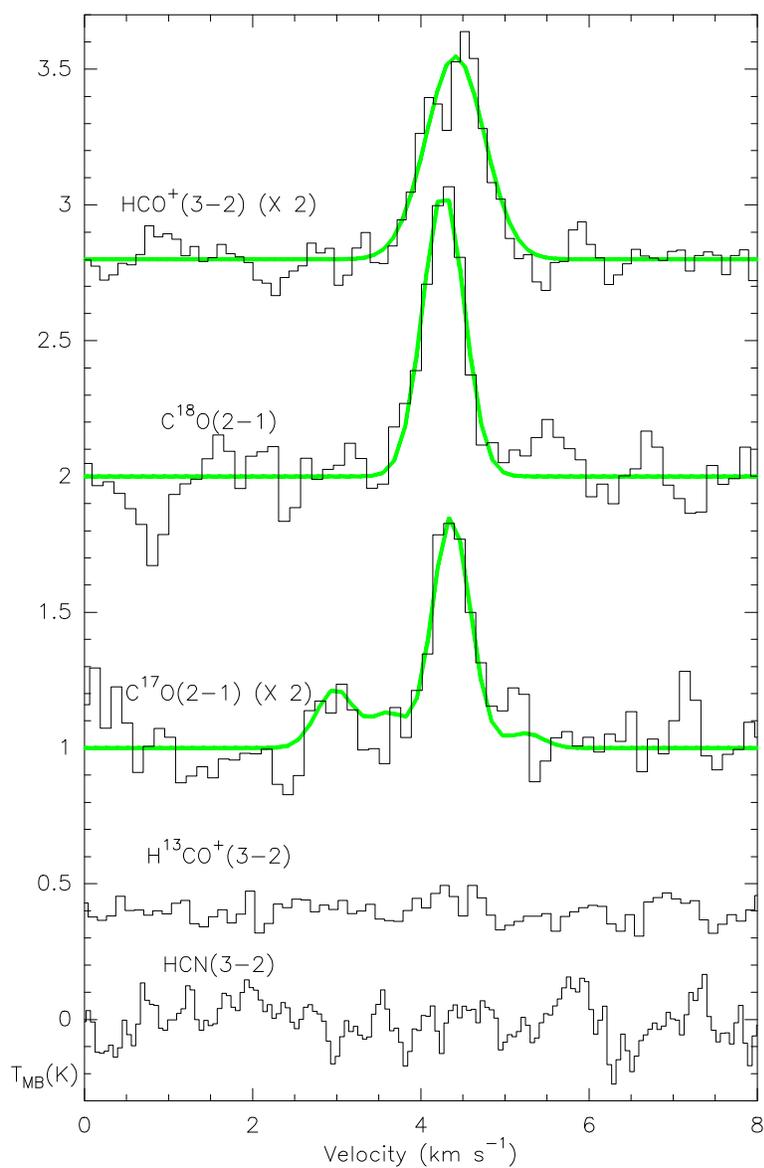}}
 \caption{\CEIO(2--1), \CSEO(2--1), \HCOP(3--2), \HTHCOP(3--2) and HCN(3--2) spectra observed 
 with the CSO towards the 1.2-mm peak of L1014. Gaussian or hyperfine fits are plotted. 
 The  asymmetric shape of \HCOP(3--2) is consistent with outward motions of the gas. \label{Fspec3}}
 \end{center}
\end{figure}

  The spectra observed towards the dust emission peak with FCRAO are shown in Figure~\ref{Fspec}
  while those observed with IRAM-30m are shown in Figure~\ref{Fspec2} and those
  taken at CSO in Figure~\ref{Fspec3}.
  Gaussian fits were performed using the CLASS fitting procedure and results are presented in
  Table~\ref{Tint}. In the case of lines with hyperfine structure, all the hyperfine components were 
  simultaneously fit. This kind of fit also furnishes the opacity of the line in the 
  presence of high S/N spectra.
  The \NTHP(1--0) line widths (0.35~\kms) are slightly broader than the prototype of
  evolved starless core L1544 \citep[0.31~\kms,][]{caselli2002a}, suggesting a combination of 
  nonthermal and systematic motions in the inner nucleus of L1014. 
  Using the CTEX approximation on the higher resolution (IRAM-30m) data, we found a column density at 
  the integrated intensity peak position 
  of $N$(\NTHP)$=6 \pm 2 \times 10^{12}$~\persc, comparable to other starless cores moderately 
  evolved in a chemical sense like e.g. L1512, L1498, L1495, L1517B \citep{crapsi2005}, and  a factor of 2-3
  less than in IRAM 04191, the other very low-luminosity Class~0 object known \citep{belloche2004}.
  Similarly, the \NTDP \ column density derived from the \NTDP(1--0) and (2--1) IRAM-30m spectra
  yielded a value of $=6 \pm 1 \times 10^{11}$~\persc \ from both lines.
  The deuterium fractionation at the peak of L1014 was, thus, found to be 10\%, a value higher
  than the average starless core \citep{crapsi2005}.
  \NTHP \ and \NTDP \ column densities were calculated also in the Large Velocity Gradient
  approximation \citep[see e.g.][for details]{crapsi2005} yielding values of $5.3 \times 10^{12}$~\persc \ 
  and   $5.2 \times 10^{11}$~\persc \ respectively. This approach furnishes also an estimate of the 
  \HH \ volume density from observations of multiple rotational transitions; in this way we
  obtained  $n$(\HH)$\simeq 1.9 \times 10^{5}$~\percc \ from \NTHP \ and  $n$(\HH)$\simeq 4.0 \times 10^{5}$~\percc \ 
  from \NTDP , both estimates are consistent within their errors with the central density evaluated 
  from the dust continuum.
    
  \TWCO (1--0) and \THCO (1--0) profiles are much more asymmetric than the other lines and their
  observed line widths  ($\sim$2~\kms \ and $\sim$1~\kms) are broader ($\sim$0.5~\kms). 
   This occurrence is not consistent with  a line  broadening caused by opacity effects alone,
   as already noted by \citet{robert1993}. 
  A possible explanation could be found considering that CO and \THCO \ emission comes from 
  the combination of emissions by randomly-moving, 
  low-density parts of the cloud with different velocities along the line of sight and with larger
  turbulence.
  \CSEO(1--0) opacity was found to be $2.4 \pm 1.5$ from hyperfine structure fitting although
  the low signal-to-noise ratio lets us think that this might be just an upper limit.
  In fact, the total column density derived from other indicators is totally inconsistent with such a
  high opacity for \CSEO .
    

  Interestingly, our CS(2--1) spectra do not show the asymmetric double peaked structure, with the blue 
  peak stronger than the red one, typical of the cores undergoing infall \citep{zhou1992,mardones1997,lee1999}. 
  This feature was  clearly seen in  IRAM 04191 \citep{andre1999}.
  Moreover, the difference in line velocity between CS and \NTHP \ in units of the \NTHP (1--0) line width
  (${\rm \delta V_{CS}= (V_{CS}-V_{N_2H^+})/\Delta V_{N_2H^+}}$), that is supposed to gauge the extended
  infall motions of the core \citep{mardones1997,lee1999}, is very small in L1014 (0.04).
  
  On the contrary, \HCOP(3--2) observations show the opposite asymmetry 
  revealing a red peak brighter than the blue peak (see Figure~\ref{Fspec3}). 
  This line profile can be produced by a rotating, expanding, or pulsating core \citep{lada2003}. 
  We note that the velocity of the self absorbed feature corresponds to
  the velocity of the optically thin \CEIO(2--1)  line as predicted by the models. 
  Given the moderate signal to noise ratio of the present data, it is necessary to confirm these 
  indications with higher sensitivity and higher resolution observations  and to retrieve a map 
  of  \HCOP \ in order to search for spatial patterns for the expansion asymmetry.
  
  We failed to detect \HTHCOP(3--2) and HCN(3--2) towards L1014-IRS 
  up to a 0.02 and 0.03~K~\kms \ level respectively. 
  
 \subsection{The search for the outflow in L1014}
\begin{figure}[htbp]
 \begin{center}
 \resizebox{9.5cm}{!}{\includegraphics{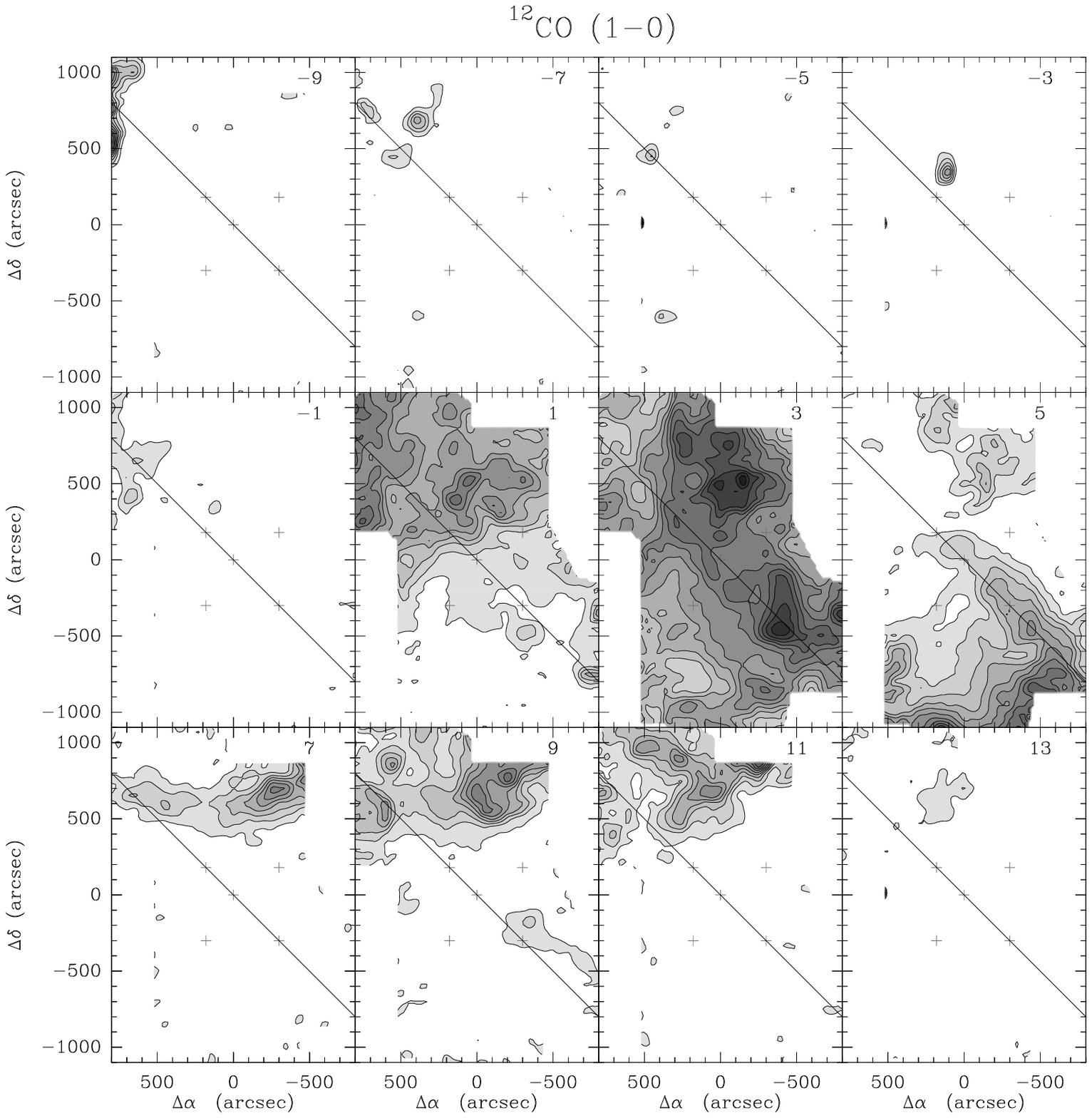}}
 \resizebox{9.5cm}{!}{\includegraphics{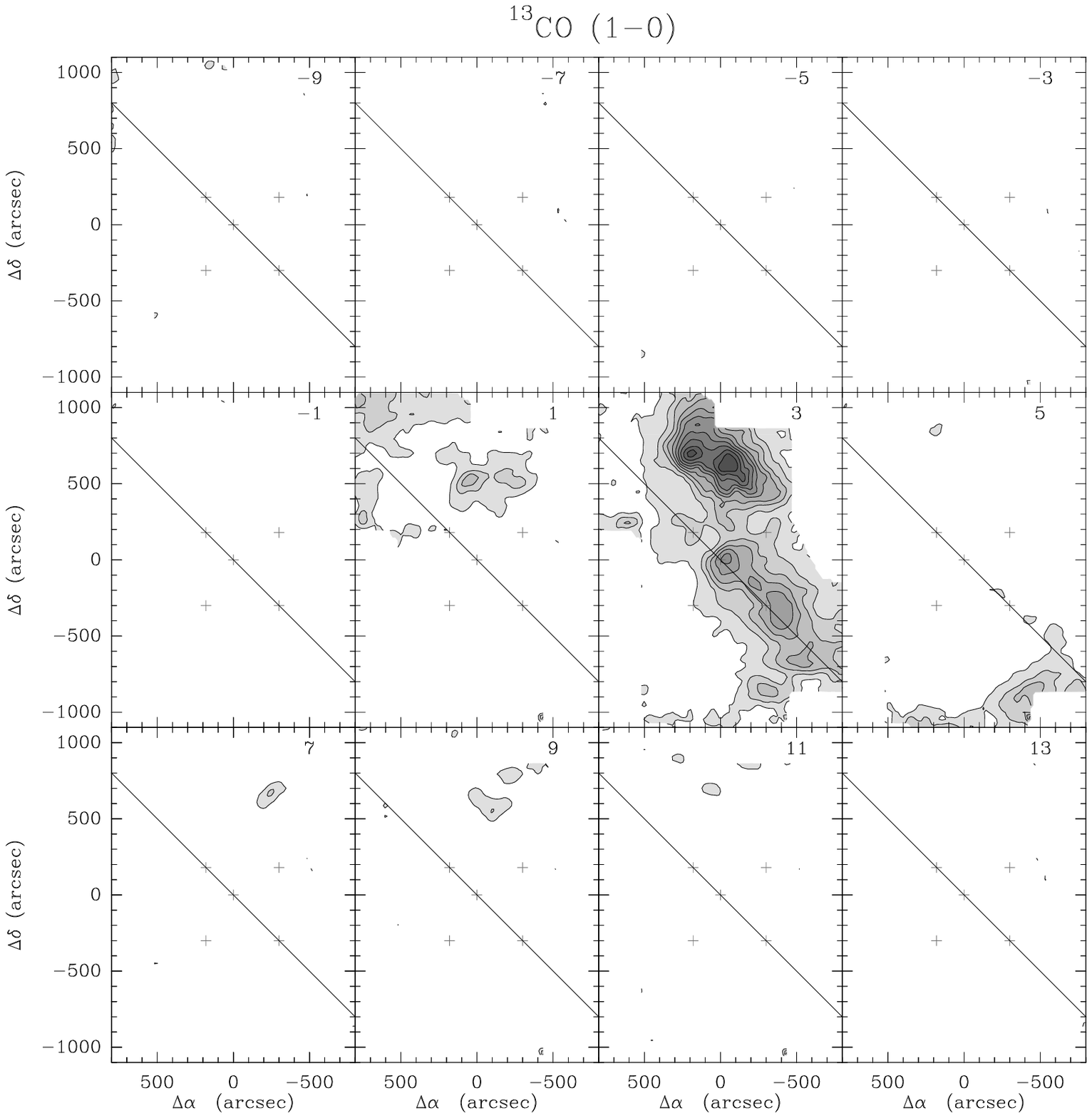}}
 \caption{ Channel maps of the CO(1--0) (top panel) and \THCO(1--0)(bottom panel) lines around
  the rest velocity of L1014.
  Contours levels start and increase by 0.8~\kks \ for the CO data and by 0.05~\kks \ for the \THCO \
  ones. Velocity bins are spaced by 2~\kms and the starting velocity of the bin is reported in
  the top right of each map. The central cross in each map indicates the dust peak while the other four 
  delineates the area shown in Figure~\ref{Fmaps}. The solid line shows the direction of the cut 
  studied in Figure~\ref{Fcut}. \label{Fco}}
 \end{center}
\end{figure}

\begin{figure}[htbp]
 \begin{center}
 \resizebox{9.5cm}{!}{\includegraphics{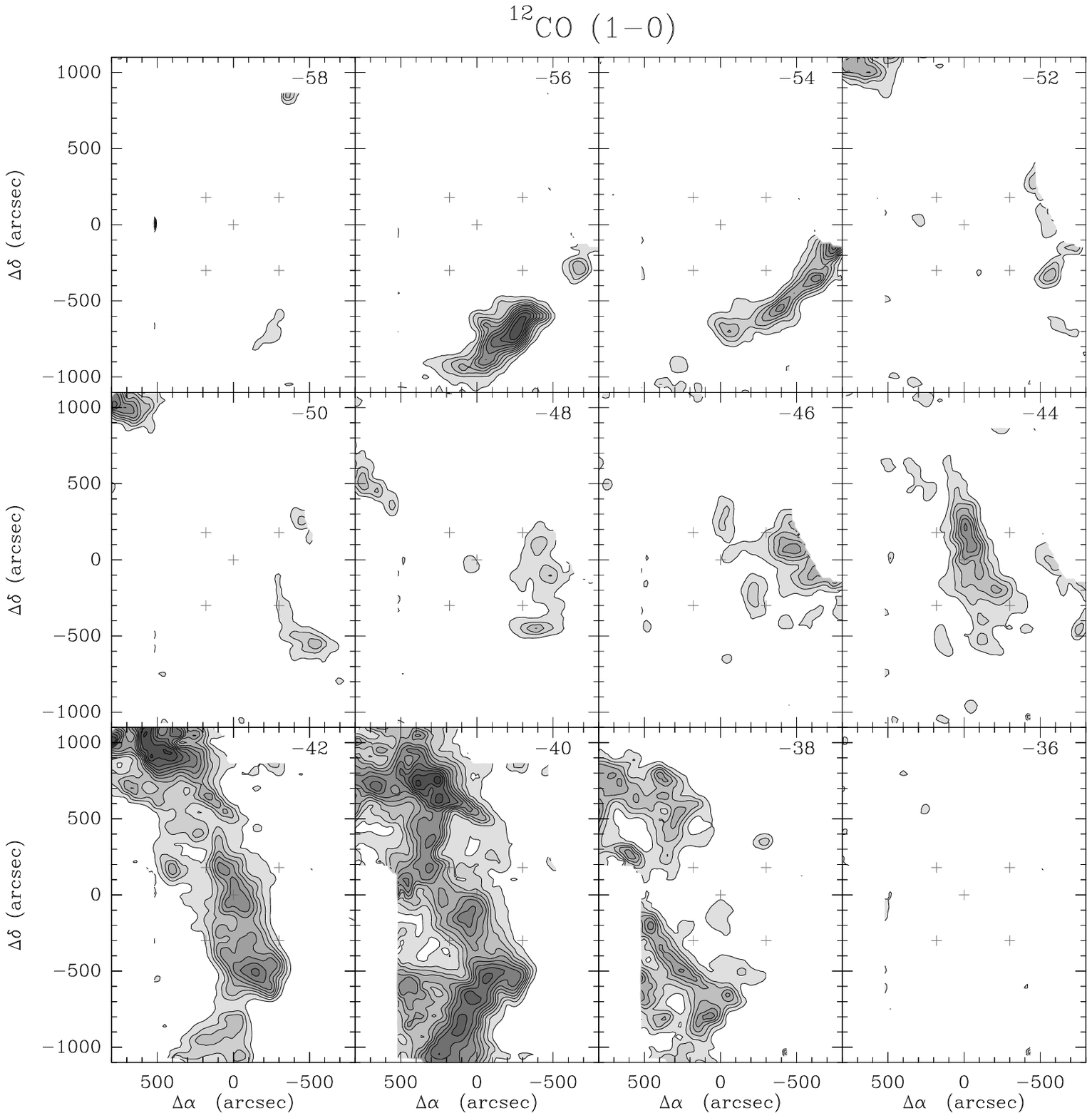}}
 \resizebox{9.5cm}{!}{\includegraphics{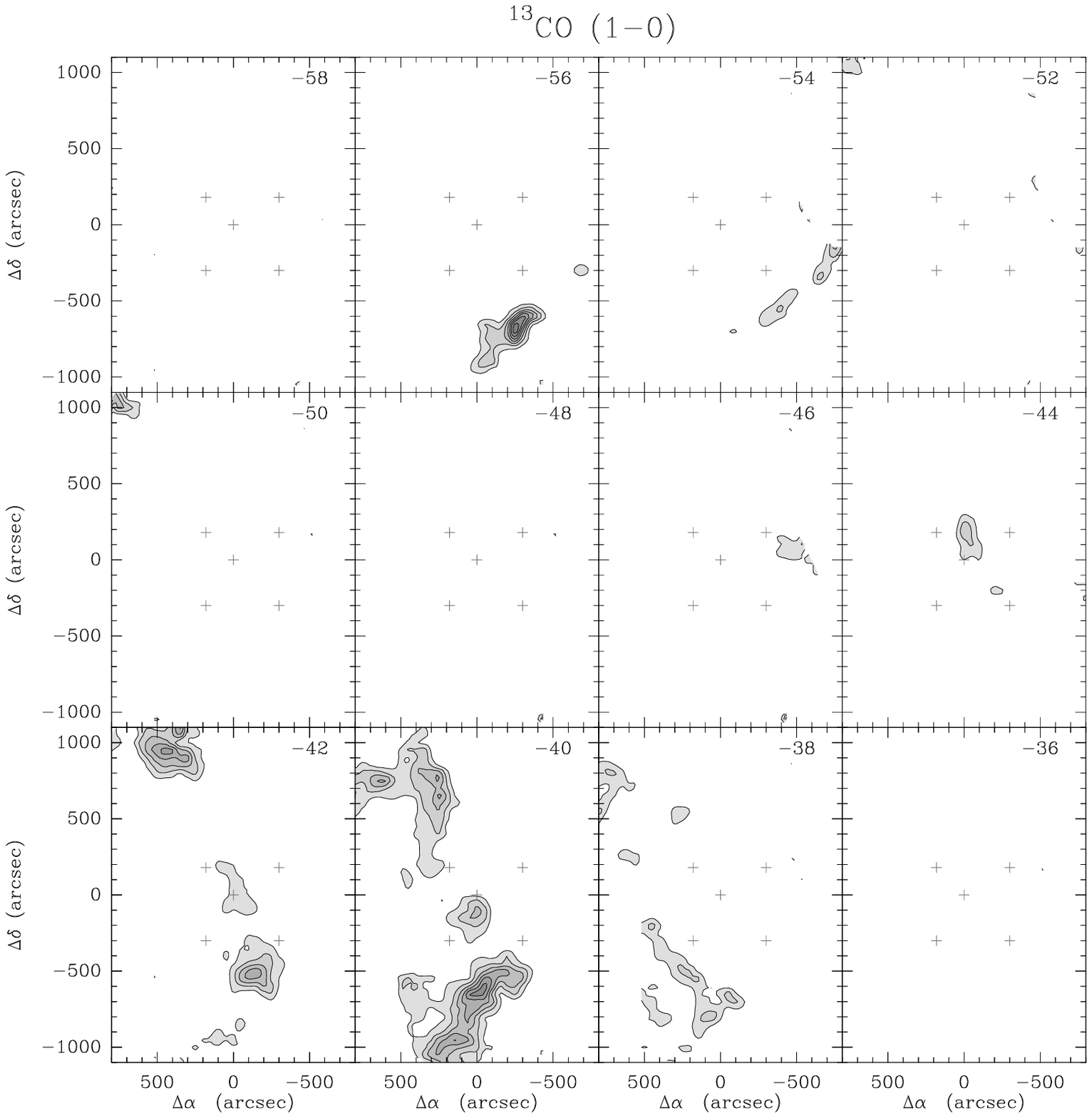}}
 \caption{Channel maps of the CO(1--0) (top panel) and \THCO(1--0)(bottom panel) lines for a second
  feature visible at $\sim-$40\kms \ and probably associated with the Perseus Arm.
  Contours levels start and increase by 0.8~\kks \ for the CO data and by 0.05~\kks \ for the \THCO \
  ones. Velocity bins are spaced by 2~\kms and the starting velocity of the bin is reported in
  the top right of each map.  The central cross in each map indicates the dust peak while the other four 
  delineates the area shown in Figure~\ref{Fmaps}.\label{Fco-40}}
 \end{center}
\end{figure}

  In Figure~\ref{Fco}, we present channel maps of CO(1--0) (top) and \THCO (1--0) (bottom) 
  in a range of velocities, $-$9~\kms \ to 15~\kms , centered around the L1014 typical velocity
  ($\sim$4.2~\kms). Emission  was integrated  in 2~\kms \ intervals.
  No CO emission was found anywhere in the mapped area between $-$35~\kms \ and $-$9~\kms \ 
  and between 15~\kms \ and 25~\kms, i.e at velocities departing more than 10~\kms \ from the L1014 
  rest velocity. The 1$\sigma$ errors for the \TWCO \ and \THCO \ channel maps are 0.06~\kks \ and
  0.03~\kks \ respectively, less than the lowest contours adopted in Figures~\ref{Fco} and \ref{Fco-40}
  
  Using IRAM 04191 \citep{andre1999} as a guide, we searched for \TWCO (1--0) wing
  emission at velocities $3 \mbox{ \kms }<|{\rm V-V_0}|<9 \mbox{ \kms }$. No wing-like emission was 
  found in this range (see maps labeled $-$5, $-$3, $-$1 and 7, 9, 11 in Figure~\ref{Fco}).
  
\begin{figure}[htbp]
 \begin{center}
 \resizebox{13cm}{!}{\includegraphics{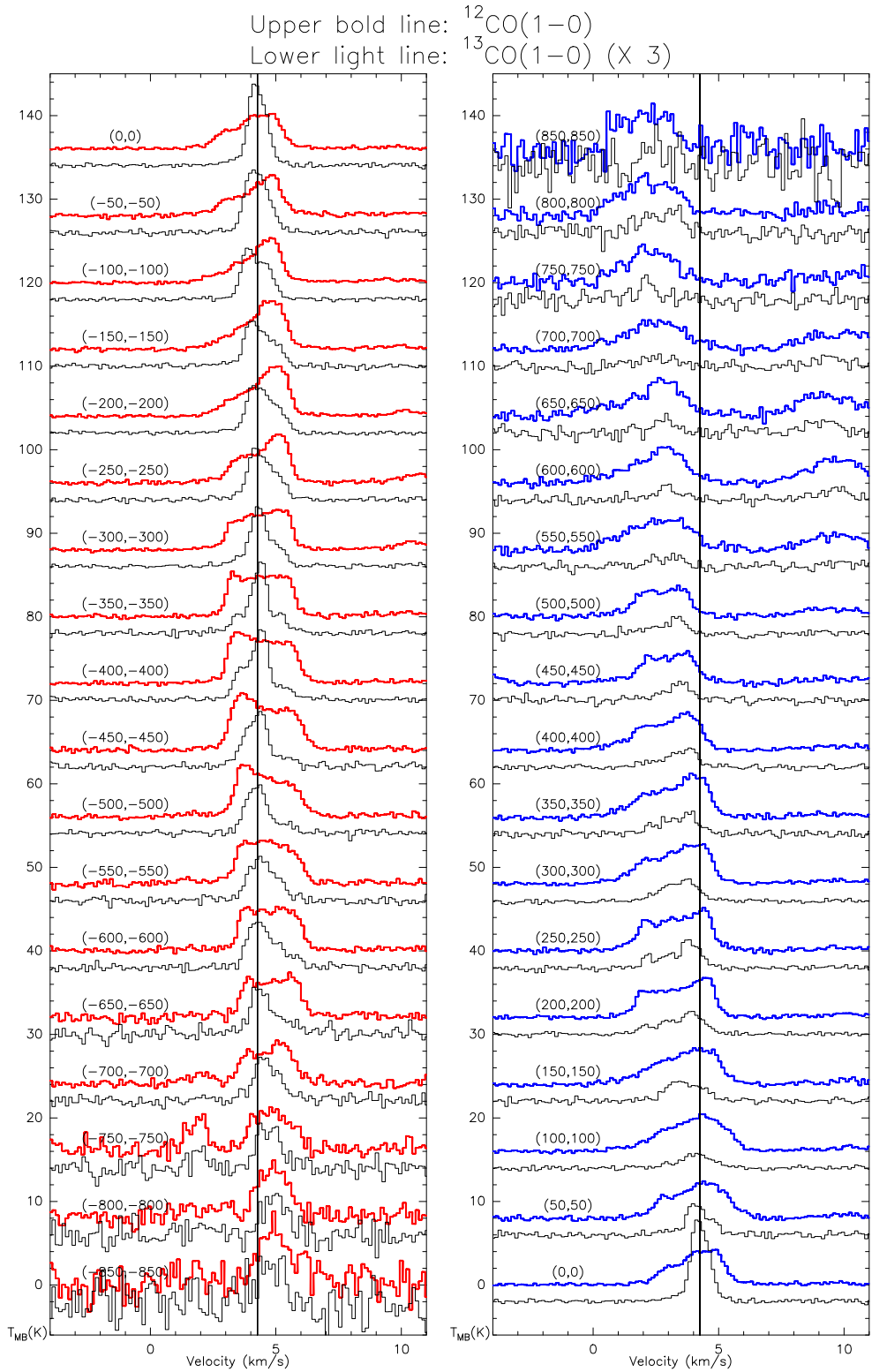}}
 \caption{CO(1--0) (upper bold line) and \THCO(1--0) (lower light line) spectra along the 
 direction of the best-candidate outflow.
 South-West (red) branch  is in the left panel and North-East (blue) branch is in the
 right panel. \THCO(1--0) spectra were multiplied by 3.\label{Fcut}}
 \end{center}
\end{figure}

  Nevertheless, given the very low luminosity of L1014-IRS we examined also the possibility of less energetic
  wings. The most likely outflow feature present in L1014 is along position angle 45\degr \  
  (measured East of North) where we can see red emission in the SW lobe between 5 and 7~\kms \ and blue 
  emission in the NE part in the bin 1 to 3~\kms. 
  This axis was identified mainly from the jet-like appearance of the ``red lobe'', but the fact
  that this feature is also seen in the \THCO (1--0) and \CEIO (1--0) emission, 
  which typically are not sensitive to outflows, poses serious questions
  as to whether this is indeed due to a classical outflow lobe.
  Moreover, as can be seen in the sequence of spectra along this axis (Figure~\ref{Fcut}), we do not see 
  the typical outflow features in the line profiles. For example:
  \begin{description}
   \item[i)]  \TWCO (1--0) line at the cloud rest velocity is not present throughout all the cut 
   and is not the dominant component in all the positions, \\
   \item[ii)] the ``low velocity wings'', seen for example in the positions ($-$500, $-$500) in the red 
   lobe or (100,100) in the blue lobe, look less like wings in the following positions ($-$200, $-$200) 
   and (400, 400).
  \end{description}
  From the spectra in Figure~\ref{Fcut}, we deduce that the red lobe and the blue extended emission
  seen in the \TWCO \  map arise from different parts of the cloud with different velocities
  along the line of sight. 
  This view is reinforced also by the observations of \THCO \ and \CEIO \ emission in the same region and
  with consistent velocities.
  The presence of gas with different velocities along the same line of sight would also explain the 
  relatively broad \TWCO \ and \THCO \ line widths found towards the entire map (see \citealt{robert1993}
  and Section~\ref{spec}).
    
  A peculiar feature of the \TWCO \ channel maps is the presence of several small spots of emission in 
  the NW part of the map at velocities smaller than 1~\kms \ and one spot at 9~\kms \ in the SE. 
  Although these may be interpreted as molecular ``bullets'' coming from the L1014 nucleus, 
  their asymmetric positioning, low velocity and large distance from the nucleus (0.05~pc) make them
  very different from the prototypical cases \citep{bachiller1996}. 
  
  We conclude that no classical outflow signature is present at the large scales investigated 
  with the FCRAO. Further sensitive observations must be performed with smaller beams to probe the CO 
  emission within the  inner 45\arcsec \  (the FCRAO beam size).

 \subsection{The background component}
        
  The Spitzer observations presented by \citet{young2004} left open the possibility
  of L1014-IRS being a more massive but less embedded distant protostar aligned by chance 
  with the L1014 nucleus.
  Although the relative rarity of embedded protostar and starless cores do not favour this  
  chance, the fact that we are looking close to the galactic plane and towards the Perseus spiral 
  arm did not allow us to disregard it just on a statistical basis. 
  Now, recent deep near-infrared observations confirm that L1014-IRS is associated
  with the nearby L1014 core (T. Huard et al. 2005, in prep.), our FCRAO 
  observations are consistent with this finding.
  
  We searched for signatures of infall-outflow activity towards L1014-IRS also at
  the  Perseus arm  velocity (near -40~\kms).  
  No  \NTHP(1--0), CS(2--1) and \CEIO(1--0) emission was detected at these velocities, most likely
  because of beam dilution given the small scales traced by those species and the large 
  Perseus arm distance \citep[2.6~kpc,][]{brand1993}.
  \citet{young2004} calculated that, assuming a distance of 2.6~kpc for L1014-IRS, the 
  protostar whose emission would fit best the observed SED would have a luminosity of 16~L$_\odot$.
  Class~I and Class~0 protostars with comparable bolometric luminosity like L1165 
  \citep{visser2002}, L1448 and L1157 \citep{bachiller1996} drive outflows of $\approx$0.3~pc in size, 
  corresponding to $\approx$20\arcsec \ at the Perseus arm distance. 
  Although this size is smaller than the FCRAO beam, we performed a search for high velocity
  wings in the $-$40~\kms \ component.
  CO and \THCO \ channel maps at these velocities are shown in Figure~\ref{Fco-40}, with the 
  same intensity scale as in  Figure~\ref{Fco}. No emission was detected below $-$58~\kms \ and above 
  $-$36~\kms .
  As for the component around 4~\kms , no clear-cut outflow features are seen in the channel map.

\section{Discussion}\label{disc}
  The contradiction between the relatively young evolutionary status of L1014 derived from chemical indicators
  and the presence of a protostellar object embedded in its nucleus is puzzling.
  L1014 shows continuum and line intensities much lower than well-studied starless cores; in particular, 
  the \HH \ volume density, \NTHP \ column density and the degree of CO depletion are smaller
  than the average starless core, from which we did not expect L1014 to be close to the star formation.
  Moreover, the CS observation does not indicate the presence of inward motions, and the 
  \HCOP \ line profile show hints of outward motions.\\  
  On the other hand, we report also relatively high deuterium fractionation and broad (if compared
  to other low-mass starless cores)  \NTHP \ and \NTDP \ lines which are typical of more evolved objects.
  
  We speculate that either i) the very low-luminosity nature of the central source 
  \citep[0.09~L$_\odot$,][]{young2004} makes its chemical and dynamical evolution different 
  to any other observed core or ii) that a low-mass,  low-luminosity ``seed'' is present long
  before the protostellar phase in every ``starless'' core.
  \citet{boss1995} and \citet{masunaga1998} modeled the early phase of protostar formation introducing the
  ``first hydrostatic core'' or ``Class~--I'' protostar as a short lived precursor of the Class~0.
  During this phase, the central temperature should reach a value of 200~K  
  (3 times lower than the temperature derived by \citealt{young2004}) and no outflow emission is expected. 
  According to the duration of this phase, we could expect the detection of similar sources in the nuclei 
  of some other starless cores in future Spitzer observations.\\

  Considering that L1014-IRS was classified as a Class~0 protostar \citep{young2004} 
  having met the requirements of $T_{bol}<70$~K and $L_{bol}/L_{smm} < 200$ \citep{andre1993}, 
  our observations indicate that L1014-IRS might be the first Class~0 protostar not associated 
  with an observed molecular outflow.  Even IRAM 04191, which has a bolometric luminosity
  comparable to L1014-IRS ($\sim$0.15~L$_\odot$), has an easily detected, very
  extended  outflow \citep{andre1999}; thus, it seems unlikely
  that the weakness of the central source is responsible for the lack of the
  outflow detection.  
  In any case, one should bear in mind that differences in the inclination angle, collimation factor,
  in the age, as well as in  the external environment could make the detection of the outflow more difficult 
  even in presence of two jets with comparable momenta.
  We can speculate that either the accretion rate of L1014-IRS is currently too little to
  power the outflow  or  that the magnetic field that threads the disk is too weak \citep{bachiller1996}.
  Alternatively, we may have caught L1014 in the epoch between the protostar formation 
  and the outflow  ignition (see e.g. \citealt{boss1995,masunaga1998}), although this hypothesis seems 
  less likely given the short lifetime associated to this phase.

  We remark that our observations do not probe small scale outflows that would be diluted in the 
  45\arcsec \ FCRAO beam, in particular if the outflow is very young and therefore compact. 
  Strong molecular outflows are a hallmark  of Class~0 protostar \citep{andre1994} , thus it is of 
  fundamental  importance to  extend the search  for the outflow to a smaller scale.
 
  The present data alone cannot rule out conclusively the  possibility of chance alignment, 
  although our search for background dense cores through \NTHP \ and CS was negative, and 
  no signatures of molecular outflows were found in the CO background component. 
  The absence of a background dense core could be explained by the chance crossing of a T-Tauri star born
  elsewhere and expelled in the direction of L1014 (see the case of PV Cephei, \citealt{goodman2004}). 
  This explanation, although difficult to rule out, is  even less probable
  than the chance alignment, given the additional requirements of a favourable trajectory
  from the parent cloud to L1014 and the coincidence in the epoch of observations.\\
  Further evidence that L1014-IRS is a young stellar object embedded within
  L1014 comes from deep near-infrared observations and will be presented by
  T. Huard et al.\ (in preparation).  

\section{Conclusions}\label{concl}
  We observed the starless core L1014 with the FCRAO antenna in \NTHP(1--0), CS(2--1), \CEIO(1--0), 
  \THCO(1--0) and \TWCO(1--0), combined with literature 1.2-mm continuum data and new \NTHP(1--0),
  \NTHP(3--2), \NTDP(1--0),  \NTDP(2--1), \NTDP(3--2) and \CSEO(1--0) observations from the IRAM-30m 
  and \CEIO(2--1), \CSEO(2--1), \HCOP(3--2), \HTHCOP(3--2) and HCN(3--2) spectra from CSO
  to study its  chemical status and to search for the presence of a molecular outflow.
  The results of our study are summarized below.
  
  1. The chemical and physical properties of L1014 derived from the present observations are not typical 
  of {\it highly evolved} low-mass starless cores. In particular, we found:
    molecular hydrogen volume density of $n$(\HH)$\simeq 2.5 \times 10^5$~\percc ,
    \NTHP \ column density of  $N$(\NTHP)$\simeq 6 \times 10^{12}$~\persc ,
    M$_{\rm vir} = 2.1$~M$_\odot$,
    CO integrated depletion factor equal to 9,
    absence of CS double peaked profile with infall signature,
    absence of velocity shifts between self-absorbed optically thick (CS) and optically thin (\NTHP) tracers
    and a profile asymmetry in  \HCOP(3--2) consistent with outward motions.
  On the other hand, we found an enhanced degree of deuterium fractionation equal to 10\% which
  is higher than the average starless core, and broad \NTHP \ and \NTDP \ lines suggestive
  of unresolved kinematical activity in the inner nucleus.
  
  These diverging indications could be reconciled considering that we are observing either a 
  very young stage of star formation or an extremely low-luminosity object exhibiting characteristics
  that differ significantly from previously known cases.

  2. No classical signatures of a molecular outflow were found towards L1014.
  In particular, we note the absence of high velocity wing and of symmetric well-defined red-blue 
  lobes in the CO channel maps.
  Bearing in mind that the scale probed by the present observations might be too large to 
  detect the outflow in L1014, we suggest that the formation of a protostar might occur prior to
  or in absence of the molecular outflow.  

  The presence of a protostar in L1014 seems to challenge the
  idea that there is a unique path to forming a protostar that all
  cores must follow.  

\begin{acknowledgements}
We gratefully thank Paola Caselli and Arnaud Belloche for taking care of the
observations at the IRAM-30m, and  the referee, Dr. Laurent  
Pagani, for clarifying several points in the manuscript.
A.C. was partly supported by NASA ``Origins of Solar System Grant'' (NAG 5-13050).
C.W.L. acknowledges supports from the Basic Research Program (KOSEF R01-2003-000-10513-0) 
of the Korea  Science and Engineering Foundation.
 This work has been partly supported by NASA "Origins of Solar
System Grant" (NNG04GG24G).
\end{acknowledgements}

\onecolumn

\begin{table}
\caption{FCRAO  settings and parameters. }
\label{Tfreq}
\centering
\begin{tabular}{lcccccc}
\hline
\hline
   line & frequency & HPBW  & T$_{SYS}$ & $\Delta v_{res}$ & vel. cov.  & Mapped region \\
   (1)  & (2)       & (3)   & (4)       & (5)              & (6) &   (7)  \\
\hline
\NTHP (1--0)   & 93.1737725 & 54 & 280 & 0.157 & $-94 \rightarrow 57$ &  5\arcmin $\times$  5\arcmin \\
CS(2--1)       & 97.980953  & 52 & 280 & 0.149 & $-90 \rightarrow 55$ &  5\arcmin $\times$  5\arcmin \\ 
\CEIO (1--0)   & 109.782173 & 46 & 280 & 0.133 & $-60 \rightarrow 70$ &  5\arcmin $\times$  5\arcmin \\ 
\THCO (1--0)   & 110.201370 & 46 & 440 & 0.133 & $-81 \rightarrow 45$ & 20\arcmin $\times$ 32\arcmin \\
\TWCO(1--0)    & 115.271203 & 44 & 920 & 0.127 & $-78 \rightarrow 43$ & 20\arcmin $\times$ 32\arcmin \\
\hline
\multicolumn{7}{l}{Note. -- Col. (2) line rest frequency~(GHz); 
Col. (3) Half Power Beam Width~(\arcsec);}\\
\multicolumn{7}{l}{Col. (4) System Temperature~(K, main beam scale);
Col. (5) Channel Spacing~(\kms);}\\
\multicolumn{7}{l}{Col. (6) Usable velocity coverage~(\kms);
Col. (7) R.A. and Dec. extension of the mapped region.}
\end{tabular}
\end{table}

\begin{table}
\caption{IRAM-30m and CSO telescope settings and parameters.} 
\label{Tfreq2}
\centering
\begin{tabular}{lcccccc}
\hline
\hline
   line & frequency & HPBW  & T$_{SYS}$ & $\Delta v_{res}$  & vel. cov.  & N$_{\rm obs}$ \\
   (1)  & (2)       & (3)   & (4)       & (5)               &   (6)      &   (7) \\
\hline
\multicolumn{7}{c}{IRAM-30m observations} \\
\hline
\NTHP (1--0)   & 93.1737725  & 26 & 165    & 0.021  & $-10 \rightarrow 16$ & 5 \\
\NTHP (3--2)   & 279.511863  &  9 & 1500   & 0.042  & $-16 \rightarrow 12$ & 3 \\
\NTDP (1--0)   &  77.109626  & 32 & 155    & 0.025  & $-14 \rightarrow 20$ & 5 \\
\NTDP (2--1)   & 154.217137  & 16 & 300    & 0.026  & $-10 \rightarrow 11$ & 3 \\
\NTDP (3--2)   & 231.321966  & 11 & 440    & 0.050  & $-16 \rightarrow 16$ & 5 \\
\CSEO (1--0)   & 112.358990  & 22 & 240    & 0.034  & $-10 \rightarrow 16$ & 1 \\ 
\hline
\multicolumn{7}{c}{CSO observations} \\
\hline
\CEIO (2--1)   & 219.560352  & 34 & 390    &  0.130 & $-29 \rightarrow 37$ & 13 \\
\CSEO (2--1)   & 224.714214  & 33 & 330    &  0.128 & $-29 \rightarrow 37$ & 5 \\
\HCOP (3--2)   & 267.557620  & 28 & 320    &  0.107 & $-21 \rightarrow 28$ & 1 \\
\HTHCOP (3--2) & 260.255478  & 29 & 300    &  0.110 & $-24 \rightarrow 32$ & 1 \\
HCN (3--2)     & 265.886434  & 28 & 500    &  0.054 & $-24 \rightarrow 31$ & 1 \\
\hline
\multicolumn{7}{l}{Note. --
Col. (2) line rest frequency~(GHz);
Col. (3) Half Power Beam Width~(\arcsec)
}\\
\multicolumn{7}{l}{
Col. (4) System Temperature~(K, main beam scale);
Col. (5) Channel Spacing~(\kms);
}\\
\multicolumn{7}{l}{
Col. (6) Usable velocity coverage~(\kms);
Col. (7) Number of observed positions.
}
\end{tabular}
\end{table}

\begin{table}
\caption{Line parameters at the 1.2-mm continuum peak position from line profile fitting. \label{Tint}}
\centering
\begin{tabular}{lcccc}
\hline
\hline
   line & intensity & $\rm V_{LSR}$  & $\Delta V$ &   $\tau$   \\
   (1)  & (2)       & (3)            & (4)        &  (5)          \\
\hline
\multicolumn{5}{c}{FCRAO spectra} \\
\hline
\NTHP(1--0)  &  0.52$\pm$0.03 &  4.239$\pm$0.014 &  0.354$\pm$0.033  &    4.6$\pm$3.2  \\
CS(2--1)     &  0.41$\pm$0.02 &  4.279$\pm$0.014 &  0.677$\pm$0.031  &    {...}	 \\
\CEIO(1--0)  &  0.84$\pm$0.02 &  4.224$\pm$0.005 &  0.484$\pm$0.012  &    {...}	 \\ 
\THCO(1--0)  &  3.18$\pm$0.07 &  4.258$\pm$0.008 &  0.864$\pm$0.019  &    {...}	 \\ 
\TWCO(1--0)  &  9.30$\pm$0.17 &  4.211$\pm$0.022 &  2.261$\pm$0.049  &    {...}	 \\ 
\hline
\multicolumn{5}{c}{IRAM-30m spectra} \\
\hline
\NTHP (1--0) &  1.850$\pm$0.032 & 4.242$\pm$0.004 & 0.354$\pm$0.009 &  6.1$\pm$0.8     \\
\NTHP (3--2) &  0.263$\pm$0.036 & 4.270$\pm$0.035 & 0.443$\pm$0.058 &    $<$0.100      \\
\NTDP (1--0) &  0.203$\pm$0.010 & 4.248$\pm$0.007 & 0.284$\pm$0.020 &  2.1$\pm$1.6     \\
\NTDP (2--1) &  0.182$\pm$0.015 & 4.273$\pm$0.010 & 0.307$\pm$0.029 &    $<$0.100      \\  
\NTDP (3--2) &  0.066$\pm$0.014 & 4.303$\pm$0.044 & 0.315$\pm$0.128 &    $<$0.100      \\
\CSEO (1--0) &  0.476$\pm$0.022 & 4.229$\pm$0.009 & 0.310$\pm$0.027 &  2.4$\pm$1.5     \\ 
\hline
\multicolumn{5}{c}{CSO spectra} \\
\hline
\CEIO (2--1)       & 0.641$\pm$0.036  & 4.270$\pm$0.019  & 0.575$\pm$0.048   & 1.4$^{a}$  \\
\CSEO (2--1)       & 0.344$\pm$0.035  & 4.092$\pm$0.023  & 0.489$\pm$0.051   & $<$ 0.100  \\
\HCOP (3--2)       & 0.316$\pm$0.015  & 4.410$\pm$0.019  & 0.819$\pm$0.041   & {...}      \\
\HTHCOP (3--2)     & 0.025$\pm$0.019  &   {...}          &  {...}            & {...}      \\
HCN (3--2)$^{c}$   &  $<$0.03         &   {...}          &  {...}            & {...}      \\  
\hline  
\multicolumn{5}{l}{Note. --
Col. (2) $\rm \int{T_{MB}dV}$~(\kks) (in presence of hyperfine structure 
}\\
\multicolumn{5}{l}{
we integrated over all the components; \TWCO \ and \THCO \ lines were 
}\\ 
\multicolumn{5}{l}{
integrated between 1 and 7 \kms \ and 3 and 5.5 \kms \ respectively); 
}\\ 
\multicolumn{5}{l}{
Col. (3) Rest velocity~(\kms)
}\\ 
\multicolumn{5}{l}{
Col. (4) Full width half maximum~(\kms);
}\\ 
\multicolumn{5}{l}{
Col. (5) Sum of the opacity of all the hyperfine components.
}\\ 
\multicolumn{5}{l}{
${a}$ Opacity was evaluated from the opacity of C$^{17}$O (2--1) and assuming the 
}\\ 
\multicolumn{5}{l}{
relative abundance as in \citet{wilson1994}.		
}\\ 
\multicolumn{5}{l}{
${c}$ Upper limit on integrated intensity was evaluated for a line-width	
}\\ 
\multicolumn{5}{l}{ equal to \HTHCOP(3--2).		
}\\ 
\end{tabular}
\end{table}

\clearpage

\begin{figure}[htbp]
 \begin{center}
 \vspace{-3cm}
 \caption{FIGURE 1 IS ATTACHED AS A JPEG FILE IN ASTRO-PF ARCHIVE.
        Optical image of the 27\arcmin$\times$37\arcmin\ field
        around L1014 from the Digital Sky Survey.  Overlaid onto the
        image are contours of beam-averaged visual extinction, following
        the {\it NICE} technique \citep[e.g.,][]{lada1994,alves1998}
        of convolving line-of-sight measurements of the H$-$K color
        excesses of 2MASS sources with a 30\arcsec\ Gaussian beam.
        The contours are drawn at A$_V$=[3, 5, 7, 9, 11] magnitudes.
        Our FCRAO observations mapped much of this region in CO(1--0)
        and \THCO(1--0), while the \NTHP(1--0), CS(2--1) and \CEIO(1--0)
        observations covered the area represented by the rectangle in
        the figure.  L1014 is seen in the center of the field, while B362
        can be seen 10\arcmin \ north of L1014.  The cross indicates the position of L1014-IRS,
        the point source observed with the Spitzer Space Telescope
        consistent with being an embedded protostar.
 \label{Fopt}}
 \end{center}
\end{figure}

\end{document}